# Strong Triplet-Exciton–LO-Phonon Coupling in Two-Dimensional Layered Organic-Inorganic Hybrid Perovskite Single Crystal Microflakes

*Yunuan Wang,[1,2] Feilong Song,[1,3] Yu Yuan,[1,3] Jianchen Dang,[1,3] Xin Xie,[1,3] Sibai Sun,[1,3] Sai Yan,[1,3] Yanbing Hou,[2] Zhidong Lou,[2,\*] and Xiulai Xu[1,3,4,#]*

[1,] Beijing National Laboratory for Condensed Matter Physics, Institute of Physics, Chinese Academy of Sciences, Beijing 100190, China

[2,] Key Laboratory of Luminescence and Optical Information, Ministry of Education, Beijing Jiaotong University, Beijing 100044, China

[3,] CAS Center for Excellence in Topological Quantum Computation and School of Physical Sciences, University of Chinese Academy of Sciences, Beijing 100049, China

[4,] Songshan Lake Materials Laboratory, Dongguan, Guangdong 523808, China

**Corresponding Author**

Zhidong Lou (zhdlou@bjtu.edu.cn)

Xiulai Xu (xlxu@iphy.ac.cn)

**Abstract**

Two-dimensional (2D) layered hybrid perovskites provide an ideal platform for studying the properties of excitons. Here, we report on a strong triplet-exciton and longitudinal-optical (LO) phonon coupling in 2D ($C_6H_5CH_2CH_2NH_3$, PEA)$_2$PbBr$_4$ perovskites. The triplet excitons exhibit strong photoluminescence (PL) in thick perovskite microflakes, and the PL is not detectable for monolayer microflakes. The coupling strength of the triplet exciton-LO phonon is approximately two to three times greater than that of the singlet exciton-LO phonon with a LO phonon energy of about 21 meV. This difference might due to the different locations of singlet excitons located in the well and triplet excitons located in the barrier in the 2D layered perovskite. Revealing the strong coupling of triplet exciton-LO phonon provides a fundamental understanding of many-body interaction in hybrid perovskites, which is useful to develop and optimize the optoelectronic devices based on 2D perovskites in the future.

Organic-inorganic hybrid perovskites have shown a great potential for applications in photovoltaic cells (PVCs),[1] lasers,[2] light-emitting diodes (LEDs),[3] and photodetectors[4] due to their superior optoelectronic properties such as large absorption coefficient,[5] long carrier diffusion length,[6] and high photoluminescence (PL) quantum yield.[7] For example, in three-dimensional (3D) perovskite based PVCs, the power conversion efficiency (PCE) has increased from 3.8% to 25.5% in about eleven years.[8,9] However, the poor humidity stability of 3D perovskites hinders the applications and commercialization of the devices. Compared with 3D perovskites, two-dimensional (2D) perovskites have higher humidity stability due to the presence of large hydrophobic organic groups.[10] Typically, 2D layered hybrid perovskites are a kind of semiconductor materials with organic and inorganic layers stacked alternately on a single molecular scale. They have a unified formula $(RNH_3)_2MX_4$, where R, M, and X is an alkyl or aromatic moiety, a metal cation, and a halide anion, respectively. Integrating different organic ammonium salts and inorganic frameworks into a 2D layered structure can achieve a variety of novel optical,[11] electrical,[12] and magnetic[13] properties. Furthermore, because of the strong quantum and dielectric confinement effects, 2D perovskites possess a large exciton binding energy,[14] which is a huge advantage for studying the characteristics of excitons at room temperature.

Exciton-phonon interaction plays a key role to understand the many-body interaction in hybrid perovskites for potential applications. Recently in $(C_6H_5CH_2CH_2NH_3, PEA)_2(CsPbBr_3)_{n-1}PbBr_4$ (n = 1-4 and ∞) perovskite thin films, the exciton-phonon coupling enhances with decreasing the number of the $[PbBr_4]^{2-}$ inorganic layer, which originates primarily from the Fröhlich interactions between excitons and longitudinal-optical (LO) phonons rather than the interactions between excitons and acoustic phonons.[15] In addition, morphology-independent white-light emission in (PEA)$_2$PbCl$_4$ has been demonstrated with strong self-trapped exciton-phonon coupling.[16] These studies mainly focused on the coupling of free excitons or self-trapped excitons with a broad emission linewidth to phonons. The coupling of triplet excitons with an ultra-narrow emission linewidth to phonons has not been investigated although a few studies on the temperature-dependent PL have been reported.[17-19] So far, researches on triplet excitons in 2D layered perovskites have been mainly focused on the influence of inorganic components with different halide anions or organic components with different numbers of carbon atoms on the formation of triplet excitons and their lifetimes.[20] For example in (C$_n$H$_{2n+1}$NH$_3$)$_2$PbBr$_4$ (n = 4, 5, 7 and 12) and (C$_6$H$_5$C$_n$H$_{2n}$NH$_3$)$_2$PbBr$_4$ (n = 1-4) perovskite thin films, the n-dependent triplet exciton emission has been confirmed with a long lifetime.[21,22] However, the triplet exciton-phonon coupling has not been investigated systematically up to now, which plays an important role in understanding the exciton properties in hybrid perovskite materials.

Different from the perovskite polycrystalline film, perovskite single crystal has less grain boundaries and lower defect states, providing a better platform to study exciton-phonon interaction in perovskites. In this work, we report on a strong triplet-exciton–LO-phonon coupling in exfoliated 2D layered (PEA)$_2$PbBr$_4$ perovskite single crystal microflakes. The layer-number-dependent triplet exciton emission is observed.

For a thick perovskite crystal microflake, the triplet excitons exhibit an ultra-narrow linewidth emission with a PL intensity much stronger than that of the singlet excitons at low temperatures. The coupling strength of the triplet exciton-LO phonon is approximately two to three times larger than that of the singlet exciton-LO phonon with a LO phonon energy of around 21 meV. The difference may be caused by the different locations of the triplet and singlet excitons with different confinement effects in the (PEA)$_2$PbBr$_4$ crystal with a natural quantum well (QW) structure. This work provides a basic understanding of the exciton-phonon interactions in 2D layered hybrid perovskites, which is important to implement potential applications in optoelectronics with the materials.

Figure 1a shows the schematic diagrams of the crystal structure and band structure of the 2D (PEA)$_2$PbBr$_4$ perovskite. The perovskite consists of a layer of inorganic framework that is constructed by co-top connected [PbBr$_4$]$^{2-}$ octahedrons and a layer of PEA$^+$ organic cations, which alternately stack in the direction perpendicular to the planes. Due to the difference between the band gap of the inorganic layer and the HOMO (highest occupied molecular orbital) – LUMO (lowest unoccupied molecular orbital) gap of the organic layer, a natural type-I QW structure with the inorganic layer as the well and the organic layer as the barrier is formed, as shown in the right panel in Figure 1a, which results in a strong quantum confinement effect in the (PEA)$_2$PbBr$_4$ perovskite. Additionally, because of the discrepancy in dielectric constant between the inorganic layer ($\varepsilon = 4.8$) and the organic layer ($\varepsilon = 3.32$),[23] the Coulomb interaction between electrons and holes in the QW can hardly be screened by the PEA$^+$ organic layer, giving rise to a strong dielectric confinement effect. Therefore, the interaction of an electron-hole pair in the well layer is enhanced by the quantum and the dielectric confinements, which results in a large oscillator strength and a large exciton binding energy of about hundreds of meV.[15,24]

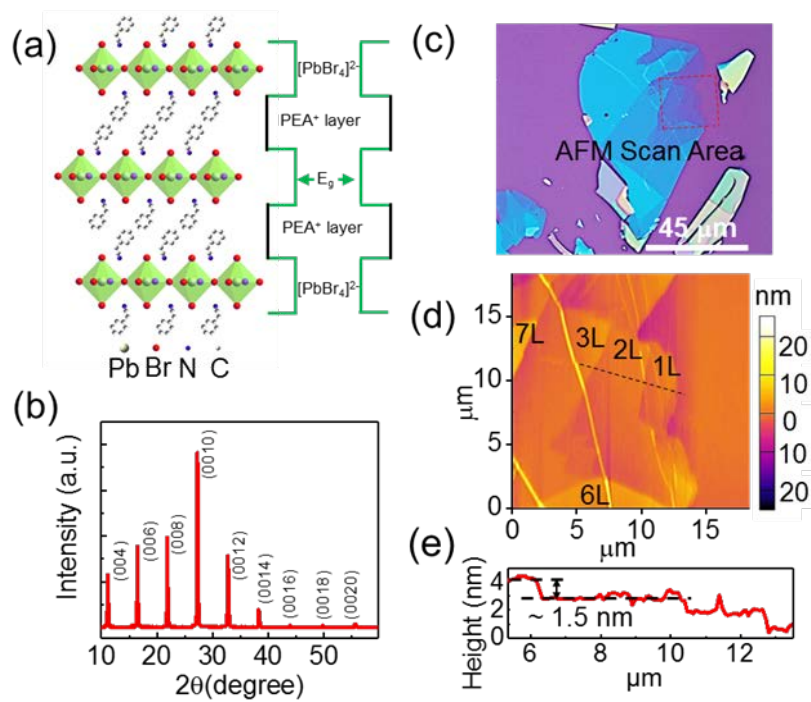

**Figure 1.** Structure and morphology of the 2D (PEA)$_2$PbBr$_4$ perovskite. (a) Schematic diagrams of the crystal structure and band structure of the hybrid perovskite. (b) X-ray diffraction (XRD) of the as-grown perovskite single crystal grown on an ITO glass substrate. (c) Optical microscope image of the single crystal microflakes with different numbers of layers on a silicon/silica substrate. (d) Atomic force microscopy (AFM) image of the thin crystal microflake with different numbers of layers (1L, 2L, 3L, 6L and 7L) corresponding to the red dashed area in (c). (e) Height profile along the black dashed line in (d).

In this work, the crystalline structure of the as-grown (PEA)$_2$PbBr$_4$ single crystal was characterized by X-ray diffraction (XRD), as shown in Figure 1b. The sharp diffraction peaks corresponding to the (00$l$, $l$ = 4, 6, 8, ..., 20) plane diffraction are observed at 11.17, 16.33, 21.81, 27.37, 32.84, 38.09, 43.96, 49.79 and 55.67°, respectively, which confirm that the organic and inorganic layers in the 2D perovskite are arranged alternately in the c-axis direction. The layered perovskites with high crystalline quality and orientation are confirmed from the equidistant angular separation.[25] An interlayer distance of about 1.51 nm between the inorganic layers is derived according to the second-order diffraction peak at 11.17° by the Bragg's formula $2d\sin\theta = n\lambda$, which is consistent with previously reported result.[26]

The exfoliated thin (PEA)$_2$PbBr$_4$ perovskite single crystal microflakes are used to study the excitonic properties of the perovskite with different numbers of layers. The morphologies and layer numbers of the exfoliated microflakes were confirmed by optical microscope and atomic force microscopy (AFM). Figure 1c exhibits an optical image of the (PEA)$_2$PbBr$_4$ microflakes, in which the perovskite microflakes with different layer numbers are clearly presented. Figure 1d shows the AFM image of the red dashed square area in Figure 1c and the height profile along the black dashed line is presented in Figure 1e. The height of the rightmost perovskite edge is ~1.5 nm, which corresponds well to the thickness of a monolayer (PEA)$_2$PbBr$_4$ perovskite determined by XRD. As the height increases by ~ 1.5 nm for each step in the profile, the number of layers of the perovskite microflake along the black dashed line in Figure 1d can be identified as one, two and three, which are labeled as 1L, 2L and 3L, respectively. At the same time, the six and seven layers marked as 6L and 7L in Figure 1d can be also determined by the height profile.

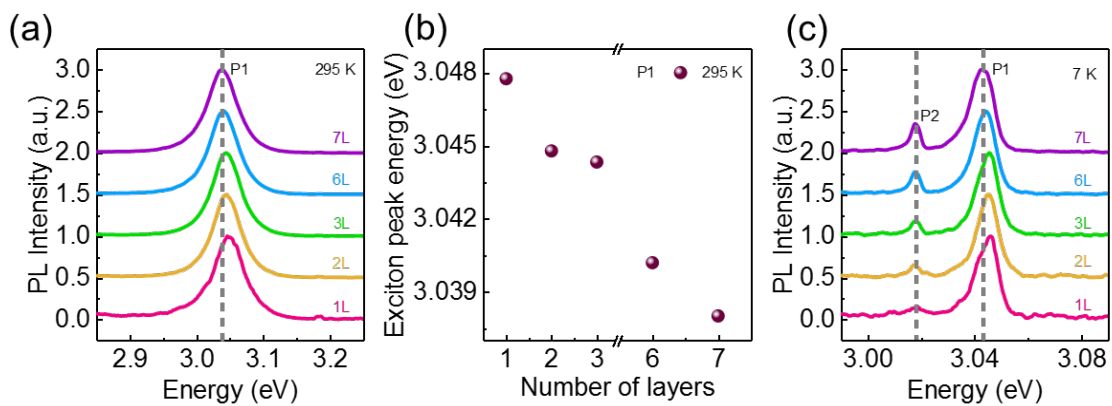

**Figure 2.** PL spectra of the thin (PEA)$_2$PbBr$_4$ perovskite single crystal micoflakes at room (295 K) and low (7 K) temperatures. (a) Normalized PL spectra of the microflakes with different numbers of layers at room temperature. For a seven-layer perovskite, a single peak (P1) at 3.038 eV is observed. The PL peak blue-shifts as the number of layers decreases. (b) The layer-number-dependent PL peak energies extracted from (a). The exciton peak energy increases by ~ 9.76 meV as the microflakes changing from seven layers to monolayer. (c) Normalized PL spectra of the microflakes with different numbers of layers at low temperature. For a seven-layer perovskite, two peaks at 3.042 (P1) and 3.018 eV (P2) are observed. The energy corresponding to P2 does not change with the number of layers. The gray dashed lines are for visual guidance only.

In order to explore the exciton features of the (PEA)$_2$PbBr$_4$ perovskite with different layer numbers, room- and low-temperature PL spectra of the exfoliated perovskite single crystal microflake with different layer numbers (1, 2, 3, 6 and 7) were investigated. The normalized PL spectra at room temperature (295 K) are shown in Figure 2a, from which we can see that there is only one exciton emission peak (P1) for the perovskite microflake with different layer numbers. For a seven-layer perovskite, an emission peak is observed at 3.038 eV (408 nm) with a full width at half maximum (FWHM) of about 53 meV. When the number of layers reduces from seven to one, the PL peak is blue-shifted, and the exciton peak energy is increased by ~ 9.76 meV, as shown in Figure 2b. The blue shift with reduced layers is similar to that of (C$_{10}$H$_{21}$NH$_3$)$_2$PbBr$_4$,[27] which can be ascribed to the increased optical band gap caused by the expansion of the lattice as the number of layers decreases.[15,28] Figure 2c displays the normalized PL spectra of the perovskite crystal microflake with different layer numbers at low temperature (7 K), the blue shift of P1 (3.042 eV) with a FWHM of 9.32 meV can still be observed clearly. However, it is worth noting that other emission peak (P2) with FWHM of 4.30 meV appears at 3.018 eV for the seven-layer perovskite, and the intensity of the PL peak gradually decreases as the number of layers decreases. For the monolayer perovskite, the PL is no more detectable. To investigate the exciton-phonon interaction, the relatively thick microflakes with strong P2 intensity are mainly discussed in the next. Different from P1, the peak energy of P2 does not shift with the number of layers. This different characteristics of the two peaks dependent on the layer number might indicate that they have different origins of radiative recombination.

Similar optical features have been observed in $(PEA)_2PbI_4$, $(C_6H_{11}NH_3)_2PbI_4$ and $(C_4H_9NH_3, BA)_2PbBr_4$ perovskites, while the origin of P2 is still under debate, which has been considered from phonon replica[24] or biexciton,[29] structural phase transition[30] and triplet exciton[18,22] etc. To determine the origin of the PL peak P2, a high-quality thick $(PEA)_2PbBr_4$ perovskite single crystal microflake (sample 1) of 1-2 μm was investigated in detail. Figure 3a depicts the power-dependent PL spectra of the sample 1 at 6 K. P1 and P2 locating at 3.041 and 3.018 eV are clearly observed when the excitation power is 400 μW, and the intensity of the peak P2 is much stronger than that of P1. Compared with the FWHM of P1 of around 5.44 meV, P2 exhibits an ultra-narrow linewidth of about 978 μeV. The narrowest linewidth in $(PEA)_2PbBr_4$ is about 839 μeV measured at an excitation power of 200 μW and at 5.3 K, as shown in the inset of Figure 3a. This value is about four times larger than the narrowest linewidth (226 µeV) of the defect state in $CH_3NH_3PbBr_3$ perovskite nanowires[31] and is the smallest for 2D perovskites so far, to our best knowledge.[18,19,32] The ultra-narrow linewidth emission of P2 suggests that the P2 peak in $(PEA)_2PbBr_4$ does not come from phonon replica, which should have a similar linewidth of the zero phonon line (P1, 5.44 meV). The dependence of the P1 and P2 intensities on the excitation power are shown in Figure 3b in double logarithmic coordinates. These two peaks exhibit a similar trend with increasing excitation power, and the coefficients of 0.98 and 0.94 for P1 and P2 are extracted respectively via fitting with a power-law function. The coefficients close to 1 are typical characteristics of single excitons, indicating that P1 and P2 derive from the single exciton states.[33] This further excludes the possibility that P2 comes from biexcitons whose dependence coefficient of PL intensity on excitation power should be around 2.[18,24,34,35]

Figure 3c displays the normalized PL spectra as a function of temperature of the sample 1. When the temperature is lower than 39.8 K, the P1 peak shows a slight blue shift with an increase in exciton peak energy by about 2.54 meV as the temperature increases. Subsequently, the P1 peak shows a red shift when the temperature is greater than 39.8 K, with an energy reduction of about 48.5 meV at room temperature. The detailed PL spectra from 5.3 to 100.0 K are illustrated in Figure 3d. No abrupt shift of the peak energy implies that there exists no phase transition in $(PEA)_2PbBr_4$ over the entire measured temperature range, which has been proved by temperature-dependent PL and Raman spectra in previous works.[20,36,37] Therefore, the P2 peak cannot be caused by the structural phase transition of the (PEA)$_2$PbBr$_4$ perovskite. In lead-based perovskite materials, the change of the bandgap with temperature causing the PL peak shifting is ascribed to the lattice expansion effect and the exciton-phonon interactions as expressed in the following formula:[15,38]

$$\frac{dE_g}{dT} = \frac{\partial E_g}{\partial V}\frac{\partial V}{\partial T} + \sum_{i,q}\left(\frac{\partial E_g}{\partial n_{i,q}}\right)\left(n_{i,q} + \frac{1}{2}\right), \quad (1)$$

where the first term is the temperature-induced lattice expansion which changes the electronic band structure by modifying the lattice constant, resulting in a bandgap blue shift with increasing temperature. $V$ represents the solid volume. The second term renders the exciton-phonon coupling that causes an electronic energy change through the lattice vibration, which in turn induces the bandgap red shifts with increasing temperature. $n_{i,q} = \frac{1}{\exp(\hbar w_{i,q}/k_B T) - 1}$, is the phonon number of the $i_{th}$ branch, where $q$, $\hbar w_{i,q}$ and $k_B$ are wave number, phonon energy and the Boltzmann constant, respectively. In this work, the P1 peak position shifting in (PEA)$_2$PbBr$_4$ with increasing temperature can be ascribed to the competition between thermal expansion effect and exciton-phonon coupling.[39] When the temperature is less than 39.8 K, the lattice expansion plays a major role and leads to the blue shift of the P1 peak position. While when the temperature is higher than the temperature, the exciton-phonon coupling dominates, resulting in a red shift of the P1 peak. Unlike the P1, the P2 peak exhibits a monotonic blue shift as the temperature increases from 5.3 K and disappears when the temperature is larger than 130 K (Figure 3c and 3d).

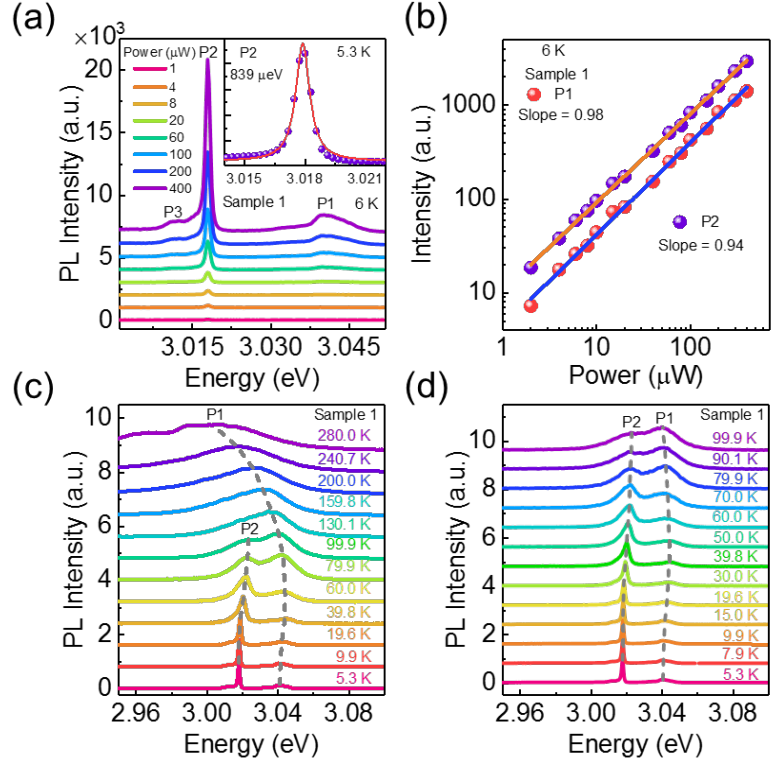

**Figure 3.** Power- and temperature-dependent PL spectra of the exfoliated thick (PEA)$_2$PbBr$_4$ perovskite single crystal micoflakes. (a) PL spectra as a function of excitation power at 6 K. The inset shows a PL spectrum with a FWHM of 839 μeV. (b) PL intensity dependent on excitation power in double logarithmic coordinates. The violet and red dots are the experimental data corresponding to P1 and P2, respectively, and the blue and orange solid lines are the fitted results with a power-law function. (c) and (d) Temperature-dependent PL spectra in temperature ranges from 5.3 to 280.0 K and 5.3 to 100.0 K. As the temperature increases, the P1 peak firstly blue shifts and then red shifts, while the P2 peak monotonously blue shifts. The gray dashed lines mark the peak position changes of P1 and P2.

Therefore, we can conclude that the peak P1 comes from the singlet exciton emission, whereas P2 comes from the triplet exciton emission, similar to those observed in (PEA)$_2$PbBr$_4$ thin film confirmed with time-resolved life-time measurements.[20,22] Given the exchange interactions, the ground states have been assigned to $\Gamma_5^-$, $\Gamma_2^-$ and $\Gamma_1^-$, where $\Gamma_5^-$ is the emission from singlet excitons (bright excitons) while $\Gamma_2^-$ and $\Gamma_1^-$ are from the triplet excitons (dark excitons).[19,23] In fact, in addition to peak P1 (3.041 eV) and P2 (3.018 eV), another relatively weak peak (P3) at 3.012 eV is also found, as shown in Figure 3a. The energy difference between P1 and P2 is 23 meV, and it is 6 meV between P2 and P3, similar to the results in (BA)$_2$PbBr$_4$ single crystals.[19] The splitting of the emission band into three fine structures corresponding to P1, P2 and P3 is caused by the exchange interactions with a short-range exchange energy $w = 32$ meV, which is similar to that of the (BA)$_2$PbBr$_4$ single crystal perovskite.[17,23] The huge exciton exchange energy is due to the small exciton Bohr radius resulting from the quantum and dielectric confinement effects in 2D perovskites with a QW structure.[23] It is known that triplet excitons are dark excitons with spin forbidden, while singlet excitons are bright excitons with spin allowed. The dark excitons (P2 and P3) are observed in (PEA)$_2$PbBr$_4$ due to the fact that the optical transitions of the dark excitons are partially allowed as a result of spin-orbit interactions and exchange interactions.[18,19] The PL intensity of triplet excitons (P2) is much higher than that of singlet excitons corresponding to P1, which is caused by the fast relaxation rate from singlet to triplet state, as demonstrated in (BA)$_2$PbBr$_4$ single crystal perovskites.[17,18] In contrast, the relatively low intensity of P2 for thin layers (as shown in Figure 2c) could be due to the different spin relaxation rates and the decay lifetimes of $\Gamma_2^-$ and $\Gamma_1^-$ for different layers.[17] The peak P3 is not discussed in detail here owing to its weak intensity.

In addition to the peak position change with increasing temperature, the linewidths of P1 and P2 peaks in the (PEA)$_2$PbBr$_4$ perovskite are broadening. The temperature-dependent PL broadening is correlated to the lattice scattering.[39] The total PL spectral broadening can be described by the following formula:[40]

$$\Gamma(T) = \Gamma_{inh} + \Gamma_{LA}T + \Gamma_{LO}\frac{1}{e^{E_{LO}/k_BT}-1} + \Gamma_{imp}e^{-E_B/k_BT} \quad , \qquad (2)$$

where the first term stands for the inhomogeneous broadening factor, and $\Gamma_{inh}$ is inhomogeneous broadening constant at 0 K; the second and third terms represent the homogeneous broadening factors, and $\Gamma_{LA}$ and $\Gamma_{LO}$ are the coupling strengths of exciton-longitudinal acoustic (LA) phonon and exciton-LO phonon, respectively. $E_{LO}$ is the LO phonon energy independent of temperature, and LO phonons are dominant in $(PEA)_2PbBr_4$ because the perovskite is a polar semiconductor.[2] The last term is an inhomogeneous broadening factor representing the scattering from ionized impurities which is governed by their average binding energy $E_B$. In organic metal halide perovskites, the scattering by ionized impurities does not play a major role, which is neglected here.[39,40] Figure 4a reveals the FWHMs of the PL peaks of sample 1 at different temperatures which were extracted from Figure 3c with a Lorentzian fit. The linewidth change with increasing temperature is also fitted with Equation 2, as shown by the solid lines in the Figure 4a. The data in the temperature range between 100 and 140 K are ignored to ensure the reliability of the fitting, owing to the spectral overlapping of P1 and P2 in this temperature range. The LO phonon energies ($E_{LO}$) of about 22.8 and 21.2 meV for the P1 and P2 peaks of sample 1 are obtained respectively. The phonons with this energy value can be attributed to the organic cation vibration, especially the π-π vibration of the phenyl groups as discussed in $(PEA)_2PbI_4$ polycrystalline thin film.[41]

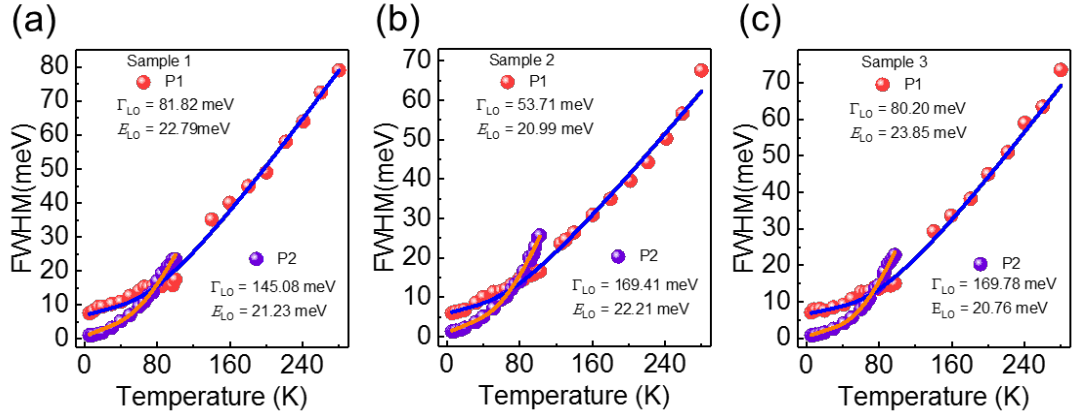

**Figure 4.** Linewidth as a function of temperature of the thick $(PEA)_2PbBr_4$ perovskite single crystal micoflakes (sample 1, 2 and 3). (a) The FWHMs of P1 and P2 peaks of sample 1 varying with temperature extracted from Figure 3c. The red and violet dots as well as the blue and orange solid lines represent the experimental data and the fitted results for P1 and P2, respectively. Similar LO phonon energies ($E_{LO}$) of about 21 meV are obtained for both P1 and P2, while the coupling intensity of the exciton-LO phonon ($\Gamma_{LO}$) of the P2 peak is about two to three times larger than that of P1 peak. (b) and (c) The FWHMs of the P1 and P2 peaks of sample 2 and 3 with increasing temperature.

Unlike the LO phonon energy, the coupling strength of the triplet exciton-LO phonon (145.08 meV) is much larger than that of singlet exciton-LO phonon (81.82 meV). This difference can be contributed to the different locations of singlet and triplet excitons in the $(PEA)_2PbBr_4$ perovskite with a QW structure. In the QW structure, the singlet excitons exist in the well, resulting in a larger quantum confinement effect than that of the triplet excitons in the barrier.[32] The confinement effect makes the exciton binding energy of the singlet excitons in $(PEA)_2PbBr_4$ (120 meV)[42] larger than the energy of the LO phonons (~21 meV), which inhibits the dissociation channel of the singlet excitons into the continuous states by absorbing LO phonon.[43] Therefore, the coupling strength of the singlet exciton-LO phonon is reduced. However, the coupling strength of the triplet exciton-LO phonon is still relatively large. This large coupling strength enhances the linewidth broadening of PL spectra of triplet excitons with temperature increase. Therefore, the emission peak of triplet excitons cannot be observed when the temperature is greater than 130 K. In order to verify the reproducibility of the experimental results, the temperature-dependent PL spectra of the other two thick $(PEA)_2PbBr_4$ perovskite micoflakes (samples 2 and 3) were measured under the same measurement conditions as sample 1. Similar results are observed as shown in Figure 4b and 4c, respectively. We can conclude that the singlet and triplet excitons are coupled with LO phonons with a similar vibration mode, but the coupling strength of the triplet exciton-LO phonon is about two to three times larger than that of the singlet exciton-LO phonon.

Two more samples (4 and 5) with monolayer perovskite micoflakes were also investigated, in which only singlet exciton emission can be observed. Figure 5a displays power-dependent PL spectra of sample 4 at low temperature. It can be seen that the intensity increases with increasing pumping power. Figure 5b shows the dependence of PL intensity on excitation laser power of sample 4 and 5. The fitted coefficients close to 1 are still obtained, which is the same as the P1 of sample 1. Figure 5c shows the temperature-dependent PL spectra of the sample 4 ranging from 10 to 280 K. The blue and red shift of the PL peak position with increasing temperature can still be clearly observed. The temperature dependence of the FWHM of the two samples are shown in Figure 5d. The PL spectral linewidth at the same temperature shows a significant difference for this two samples, which might due to the different crystalline quality. By fitting the results with Equation 2, a LO phonon energy of about 21 meV can be obtained with a relatively small coupling strength, which is similar to the singlet exciton in thick (PEA)$_2$PbBr$_4$ perovskite single crystal micoflakes.

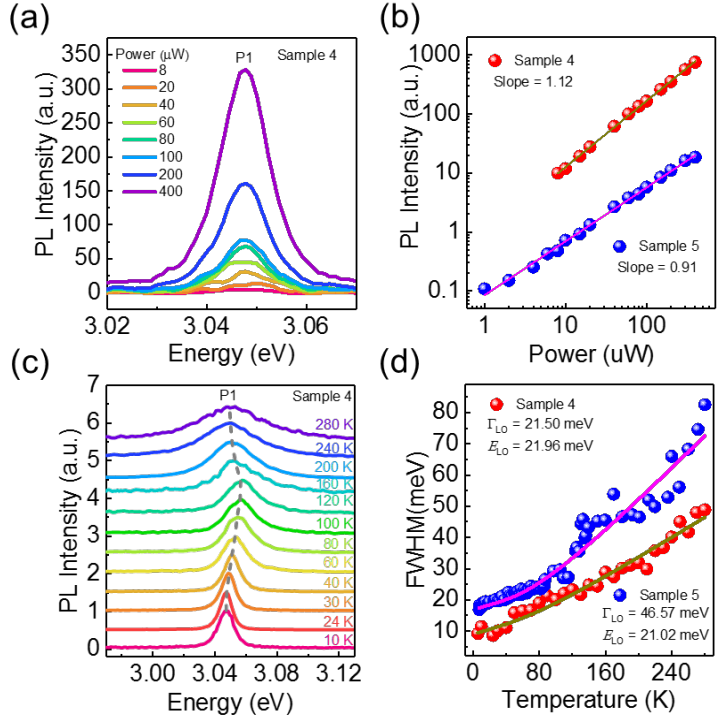

**Figure 5.** Power- and temperature-dependent PL spectra for monolayer perovskite micoflakes. (a) PL spectra with different excitation light powers at 5.8 K from sample 4, a monolayer (PEA)$_2$PbBr$_4$ perovskite micoflake. Only one emission peak (P1) was observed. (b) The intensity of the PL spectra as a function of the excitation light power in sample 4 and 5 monolayer perovskite micoflakes. The dots and solid lines represent the experimental data and fitted results. The fitted coefficients of samples 4, 5 are 1.12 and 0.91, respectively. (c) The variations of PL spectra with temperature for sample 4. The gray dotted line shows the change of the P1 peak position with increasing temperature. (d) The FWHMs of the PL peaks as a function of temperature for sample 4 and 5. The solid lines show the fitted results.

All the fitted results including exciton-LA phonon coupling are summarized in Table 1. The exciton-LA phonon coupling strength is less than 0.1 indicating that the broadening of the PL spectra by acoustic phonons is weak, which shows that the exciton-LO phonon Fröhlich interactions dominates the exciton-phonon coupling. The exciton-LO phonon interactions are the main reason for the broadening of the PL spectra regardless of singlet excitons or triplet excitons in 2D (PEA)$_2$PbBr$_4$ perovskite single crystal micoflakes. The coupling strength of the triplet exciton-LO phonon is about two to three times greater than that of the singlet exciton-LO phonon for the same sample.

**Table 1.** Fitted results of FWHM with temperature for the perovskite single crystal microflakes with different numbers of layers. Where $\Gamma_{inh}$, $\Gamma_{LA}$, $\Gamma_{LO}$ and $E_{LO}$ are inhomogeneous broadening constant at 0 K, the coupling strength of exciton-longitudinal acoustic phonon (LA), the coupling strength of exciton-longitudinal optical phonon (LO) and LO phonon energy, respectively.

|          | Peak | $\Gamma_{inh}$ [meV] | $\Gamma_{LA}$ [meV K$^{-1}$] | $\Gamma_{LO}$ [meV] | $E_{LO}$ [meV] |
|----------|------|------|------|--------|-------|
| Sample 1 | P1   | 6.97 | 0.07 | 81.82  | 22.79 |
|          | P2   | 0.82 | 0.10 | 145.08 | 21.23 |
| Sample 2 | P1   | 5.89 | 0.06 | 53.71  | 20.99 |
|          | P2   | 1.04 | 0.09 | 169.41 | 22.21 |
| Sample 3 | P1   | 6.65 | 0.05 | 80.20  | 23.85 |
|          | P2   | 0.56 | 0.08 | 169.78 | 20.76 |
| Sample 4 | P1   | 8.94 | 0.08 | 21.50  | 21.96 |
| Sample 5 | P1   | 16.83| 0.08 | 46.57  | 21.02 |

In summary, different numbers of layers of 2D (PEA)$_2$PbBr$_4$ single crystal microflakes have been obtained by mechanically exfoliating from high-quality 2D perovskite bulk single crystals. The PL from triplet exciton emission with a narrow linewidth about 839 μeV has been observed at low temperature from a thick (PEA)$_2$PbBr$_4$ perovskite micoflake. As the layer number decreases, the PL intensity from triplet exciton emission decreases and not detectable in monolayer. Furthermore, we found that the coupling strength of the triplet exciton-LO phonon is two to three times greater than that of the singlet exciton-LO phonon with a phonon energy of approximately 21 meV. The strong triplet exciton-phonon interaction might due to the fact that triplet excitons locate at the interface between organic and inorganic in 2D perovskites, while the small singlet exciton-phonon interaction is due to the confinement of the QW structure. Evidence of strong coupling between triplet exciton-LO phonon provides a basis for exploring triplet excitons for future applications in developing high-performance optoelectronic devices.

**Experimental Methods**

The as-grown 2D layered $(PEA)_2PbBr_4$ perovskite bulk crystal was synthesized using an anti-solvent vapor-assisted method as shown in our previous work.[10] The perovskite single crystal microflakes with different layer numbers were obtained by mechanical exfoliation with a Scotch tape from the as-grown perovskite crystal. The exfoliated perovskite microflakes with different layer numbers attached to the tape were transferred to a silicon substrate covered with 300-nm silicon dioxide using polydimethylsiloxane (PDMS). The XRD spectrum was recorded by a Bruker D8 X-ray diffractometer using Cu Kα radiation with a wavelength of 1.54 Å. AFM measurements were executed with an AC Air Mode in the ambient condition. Both power-dependent and temperature-dependent PL spectra were obtained using a continuous flow liquid helium cooling system with an excitation wavelength at 325 nm. The laser was focused on the perovskite crystal microflakes by a reflective microscope objective lens with a spot radius of around 1 - 2 μm. The PL signal from the sample was dispersed by a grating spectrometer and collected by a liquid-nitrogen-cooled charge-coupled device camera with a spectral resolution of 60 μeV.

**Acknowledgements**

This work was supported by the National Natural Science Foundation of China (Grants Nos. 62025507, 11934019, 11721404, 11874419 and 62075009), the Key-Area Research and Development Program of Guangdong Province (Grant No.2018B030329001), and the Strategic Priority Research Program (Grant No. XDB28000000), the Instrument Developing Project (Grant No. YJKYYQ20180036) and the Interdisciplinary Innovation Team of the Chinese Academy of Science.

**References**

(1) Jeong, M.; Choi, I. W.; Go, E. M.; Cho, Y.; Kim, M.; Lee, B.; Jeong, S.; Jo, Y.; Choi, H. W.; Lee, J.; et al. Stable perovskite solar cells with efficiency exceeding 24.8% and 0.3-V voltage loss. *Science* **2020,** *369*, 1615-1620.

(2) Liang, Y.; Shang, Q.; Wei, Q.; Zhao, L.; Liu, Z.; Shi, J.; Zhong, Y.; Chen, J.; Gao, Y.; Li, M.; et al. Lasing from Mechanically Exfoliated 2D Homologous Ruddlesden-Popper Perovskite Engineered by Inorganic Layer Thickness. *Adv. Mater.* **2019,** *31*, 1903030.

(3) Zhang, C.; Wang, S.; Li, X.; Yuan, M.; Turyanska, L.; Yang, X. Core/Shell Perovskite Nanocrystals: Synthesis of Highly Efficient and Environmentally Stable $FAPbBr_3$ /$CsPbBr_3$ for LED Applications. *Adv. Funct. Mater.* **2020,** *30*, 1910582.

(4) Liu, Y.; Ye, H.; Zhang, Y.; Zhao, K.; Yang, Z.; Yuan, Y.; Wu, H.; Zhao, G.; Yang, Z.; Tang, J.; et al. Surface-Tension-Controlled Crystallization for High-Quality 2D Perovskite Single Crystals for Ultrahigh Photodetection. *Matter* **2019,** *1*, 465-480.

(5) Kwon, U.; Hasan, M. M.; Yin, W.; Kim, D.; Ha, N. Y.; Lee, S.; Ahn, T. K.; Park, H. J. Investigation into the Advantages of Pure Perovskite Film without $PbI_2$ for High Performance Solar Cell. *Sci. Rep.* **2016,** *6*, 35994.

(6) Dong, Q.; Fang, Y.; Shao, Y.; Mulligan, P.; Qiu, J.; Cao, L.; Huang, J. Electron-hole diffusion lengths > 175 μm in solution-grown $CH_3NH_3PbI_3$ single crystals. *Science* **2015,** *347*, 967-970.

(7) Yuan, M.; Quan, L. N.; Comin, R.; Walters, G.; Sabatini, R.; Voznyy, O.; Hoogland, S.; Zhao, Y.; Beauregard, E. M.; Kanjanaboos, P.; et al. Perovskite energy funnels for efficient light-emitting diodes. *Nat. Nanotechnol.* **2016,** *11*, 872-877.

(8) Kojima, A.; Teshima, K.; Shirai, Y.; Miyasaka, T. Organometal Halide Perovskites as Visible-Light Sensitizers for Photovoltaic Cells. *J. Am. Chem. Soc.* **2009,** *131*, 6050–6051.

(9) See the most recent certified efficiency at https://www.nrel.gov/pv/cell-efficiency.html.

(10) Wang, Y.; Tang, Y.; Jiang, J.; Zhang, Q.; Sun, J.; Hu, Y.; Cui, Q.; Teng, F.; Lou, Z.; Hou, Y. Mixed-dimensional self-assembly organic–inorganic perovskite microcrystals for stable and efficient photodetectors. *J. Mater. Chem. C* **2020,** *8*, 5399-5408.

(11) Ma, J.; Fang, C.; Chen, C.; Jin, L.; Wang, J.; Wang, S.; Tang, J.; Li, D. Chiral 2D Perovskites with a High Degree of Circularly Polarized Photoluminescence. *ACS Nano* **2019,** *13*, 3659-3665.

(12) Leng, K.; Abdelwahab, I.; Verzhbitskiy, I.; Telychko, M.; Chu, L.; Fu, W.; Chi, X.; Guo, N.; Chen, Z.; Chen, Z.; et al. Molecularly thin two-dimensional hybrid perovskites with tunable optoelectronic properties due to reversible surface relaxation. *Nat. Mater.* **2018,** *17*, 908-914.

(13) Long, G.; Jiang, C.; Sabatini, R.; Yang, Z.; Wei, M.; Quan, L. N.; Liang, Q.; Rasmita, A.; Askerka, M.; Walters, G.; et al. Spin control in reduced-dimensional chiral perovskites. *Nat. Photonics* **2018,** *12*, 528-533.

(14) Shang, Q.; Wang, Y.; Zhong, Y.; Mi, Y.; Qin, L.; Zhao, Y.; Qiu, X.; Liu, X.; Zhang, Q. Unveiling Structurally Engineered Carrier Dynamics in Hybrid Quasi-Two-Dimensional Perovskite Thin Films toward Controllable Emission. *J. Phys. Chem. Lett.* **2017,** *8*, 4431-4438.

(15) Long, H.; Peng, X.; Lu, J.; Lin, K.; Xie, L.; Zhang, B.; Ying, L.; Wei, Z. Exciton-phonon interaction in quasi-two dimensional layered $(PEA)_2(CsPbBr_3)_{n-1}PbBr_4$ perovskite. *Nanoscale* **2019,** *11*, 21867-21871.

(16) Thirumal, K.; Chong, W. K.; Xie, W.; Ganguly, R.; Muduli, S. K.; Sherburne, M.; Asta, M.; Mhaisalkar, S.; Sum, T. C.; Soo, H. S.; et al. Morphology-Independent Stable White-Light Emission from Self-Assembled Two-Dimensional Perovskites Driven by Strong Exciton–Phonon Coupling to the Organic Framework. *Chem. Mater.* **2017,** *29*, 3947-3953.

(17) Ema, K.; Umeda, K.; Toda, M.; Yajima, C.; Arai, Y.; Kunugita, H.; Wolverson, D.; Davies, J. J. Huge exchange energy and fine structure of excitons in an organic-inorganic quantum well material. *Phys. Rev. B* **2006,** *73*, 241310.

(18) Yamamoto, Y.; Oohata, G.; Mizoguchi, K.; Ichida, H.; Kanematsu, Y. Photoluminescence of excitons and biexcitons in $(C_4H_9NH_3)_2PbBr_4$ crystals under high excitation density. *Phys. Status Solidi C* **2012,** *9*, 2501-2504.

(19) Tanaka, K.; Takahashi, T.; Kondo, T.; Umeda, K.; Ema, K.; Umebayashi, T.; Asai, K.; Uchida, K.; Miura, N. Electronic and Excitonic Structures of Inorganic–Organic Perovskite-Type Quantum-Well Crystal $(C_4H_9NH_3)_2PbBr_4$. *Jpn. J. Appl. Phys.* **2005,** *44*, 5923-5932.

(20) Kitazawa, N.; Aono, M.; Watanabe, Y. Temperature-dependent time-resolved photoluminescence of $(C_6H_5C_2H_4NH_3)_2PbX_4$ (X=Br and I). *Mater. Chem. Phys.* **2012,** *134*, 875-880.

(21) Kitazawa, N.; Aono, M.; Watanabe, Y. Excitons in organic–inorganic hybrid compounds $(C_nH_{2n+1}NH_3)_2PbBr_4$ (n=4, 5, 7 and 12). *Thin Solid Films* **2010,** *518*, 3199-3203.

(22) Kitazawa, N.; Watanabe, Y. Optical properties of natural quantum-well compounds $(C_6H_5-C_nH_{2n}-NH_3)_2PbBr_4$ (n=1–4). *J. Phys. Chem. Solids* **2010,** *71*, 797-802.

(23) Takagi, H.; Kunugita, H.; Ema, K. Influence of the image charge effect on excitonic energy structure in organic-inorganic multiple quantum well crystals. *Phys. Rev. B* **2013,** *87*, 125421.

(24) Gauthron, K.; Lauret, J.-S.; Doyennette, L.; Lanty, G.; Al Choueiry, A.; Zhang, S.; Brehier, A.; Largeau, L.; Mauguin, O.; Bloch, J.; et al. Optical spectroscopy of two-dimensional layered $(C_6H_5C_2H_4-NH_3)_2-PbI_4$ perovskite. *Opt. Express* **2010,** *18*, 5912-5919.

(25) Cao, D. H.; Stoumpos, C. C.; Farha, O. K.; Hupp, J. T.; Kanatzidis, M. G. 2D Homologous Perovskites as Light-Absorbing Materials for Solar Cell Applications. *J.*

*Am. Chem. Soc.* **2015,** *137*, 7843-7850.

(26) Tabuchi, Y.; Asai, K.; Rikukawa, M.; Sanui, K.; Ishigure, K. Preparation and characterization of natural lower dimensional layered perovskite-type compounds. *J. Phys. Chem. Solids* **2000,** *61*, 837–845.

(27) Yin, H.; Jin, L.; Qian, Y.; Li, X.; Wu, Y.; Bowen, M. S.; Kaan, D.; He, C.; Wozniak, D. I.; Xu, B.; et al. Excitonic and Confinement Effects of 2D Layered $(C_{10}H_{21}NH_3)_2PbBr_4$ Single Crystals. *ACS Appl. Energy Mater.* **2018,** *1*, 1476-1482.

(28) Dou, L.; Wong, A. B.; Yu, Y.; Lai, M.; Kornienko, N.; Eaton, S. W.; Fu, A.; Bischak, C. G.; Ma, J.; Ding, T.; et al. Atomically thin two-dimensional organic-inorganic hybrid perovskites. *Science* **2015,** *349*, 1518-1521.

(29) Fujisawa, J.-i.; Ishihara, T. Excitons and biexcitons bound to a positive ion in a bismuth-doped inorganic-organic layered lead iodide semiconductor. *Phys. Rev. B* **2004,** *70*, 205330.

(30) Yangui, A.; Pillet, S.; Mlayah, A.; Lusson, A.; Bouchez, G.; Triki, S.; Abid, Y.; Boukheddaden, K. Structural phase transition causing anomalous photoluminescence behavior in perovskite $(C_6H_{11}NH_3)_2[PbI_4]$. *J. Chem. Phys.* **2015,** *143*, 224201.

(31) Song, F.; Qian, C.; Wang, Y.; Zhang, F.; Peng, K.; Wu, S.; Xie, X.; Yang, J.; Sun, S.; Yu, Y.; et al. Hot Polarons with Trapped Excitons and Octahedra‐Twist Phonons in $CH_3NH_3PbBr_3$ Hybrid Perovskite Nanowires. *Laser Photonics Rev.* **2020,** *14*, 1900267.

(32) Goto, T.; Makino, H.; Yao, T.; Chia, C. H.; Makino, T.; Segawa, Y.; Mousdis, G. A.; Papavassiliou, G. C. Localization of triplet excitons and biexcitons in the two-dimensional semiconductor $(CH_3C_6H_4CH_2NH_3)_2PbBr_4$. *Phys. Rev. B* **2006,** *73*, 115206.

(33) Yu, Y.; Dang, J.; Qian, C.; Sun, S.; Peng, K.; Xie, X.; Wu, S.; Song, F.; Yang, J.; Xiao, S.; et al. Many-body effect of mesoscopic localized states in $MoS_2$ monolayer. *Phys. Rev. Mater.* **2019,** *3*, 051001.

(34) Qian, C.; Wu, S.; Song, F.; Peng, K.; Xie, X.; Yang, J.; Xiao, S.; Steer, M. J.; Thayne, I. G.; Tang, C.; et al. Two-Photon Rabi Splitting in a Coupled System of a Nanocavity and Exciton Complexes. *Phys. Rev. Lett.* **2018,** *120*, 213901.

(35) Peng, K.; Wu, S.; Tang, J.; Song, F.; Qian, C.; Sun, S.; Xiao, S.; Wang, M.; Ali, H.; Williams, D. A.; et al. Probing the Dark-Exciton States of a Single Quantum Dot Using Photocurrent Spectroscopy in a Magnetic Field. *Phys. Rev. Appl.* **2017,** *8*, 064018.

(36) Gong, X.; Voznyy, O.; Jain, A.; Liu, W.; Sabatini, R.; Piontkowski, Z.; Walters, G.; Bappi, G.; Nokhrin, S.; Bushuyev, O.; et al. Electron-phonon interaction in efficient perovskite blue emitters. *Nat. Mater.* **2018,** *17*, 550-556.

(37) Dhanabalan, B.; Leng, Y.-C.; Biffi, G.; Lin, M.-L.; Tan, P.-H.; Infante, I.; Manna, L.; Arciniegas, M. P.; Krahne, R. Directional Anisotropy of the Vibrational Modes in 2D-Layered Perovskites. *ACS Nano* **2020,** *14*, 4689-4697.

(38) Yu, C.; Chen, Z.; Wang, J. J.; Pfenninger, W.; Vockic, N.; Kenney, J. T.; Shum, K. Temperature dependence of the band gap of perovskite semiconductor compound $CsSnI_3$. *J. Appl. Phys.* **2011,** *110*, 063526.

(39) Peng, S.; Wei, Q.; Wang, B.; Zhang, Z.; Yang, H.; Pang, G.; Wang, K.; Xing, G.; Sun, X. W.; Tang, Z. Suppressing Strong Exciton-Phonon Coupling in Blue Perovskite Nanoplatelet Solids by Binary Systems. *Angew.Chem. Int. Ed.* **2020,** *59*, 22156-22162.

(40) Wright, A. D.; Verdi, C.; Milot, R. L.; Eperon, G. E.; Perez-Osorio, M. A.; Snaith, H. J.; Giustino, F.; Johnston, M. B.; Herz, L. M. Electron-phonon coupling in hybrid lead halide perovskites. *Nat. Commun.* **2016,** *7*, 11755.

(41) Thouin, F.; Cortecchia, D.; Petrozza, A.; Srimath, K. A. R.; Silva, C. Enhanced screening and spectral diversity in many-body elastic scattering of excitons in two-dimensional hybrid metal-halide perovskites. *Phys. Rev. Res.* **2019,** *1*, 032032.

(42) Zhai, W.; Ge, C.; Fang, X.; Zhang, K.; Tian, C.; Yuan, K.; Sun, S.; Li, Y.; Chen, W.; Ran, G. Acetone vapour-assisted growth of 2D single-crystalline organic lead halide perovskite microplates and their temperature-enhanced photoluminescence. *RSC Adv.* **2018,** *8*, 14527-14531.

(43) Pelekanos, N. T.; Ding, J.; Hagerott, M.; Nurmikko, A. V.; Luo, H.; Samarth, N.; Furdyna, J. K. Quasi-two-dimensional excitons in (Zn,Cd)Se/ZnSe quantum wells: Reduced exciton-LO-phonon coupling due to confinement effects. *Phys. Rev. B* **1992,** *45*, 6037.